\documentclass[conference]{IEEEtran}
\IEEEoverridecommandlockouts
\usepackage{cite}
\usepackage{amsmath}
\usepackage{multirow} 
\usepackage{multicol} 
\usepackage{arydshln}
\usepackage{graphicx}
\usepackage{makecell}
\usepackage{colortbl}
\usepackage{xcolor}
\usepackage{array}
\newcolumntype{C}[1]{>{\centering\arraybackslash}p{#1}}
\usepackage{indentfirst}
\usepackage{booktabs}
\usepackage{color}
\usepackage{url}
\definecolor{Green}{RGB}{92,182,64}
\begin{document}

\title{SEDAC: A CVAE-Based Data Augmentation Method for Security Bug Report Identification}

\author{\IEEEauthorblockN{Yansheng Liao}
\IEEEauthorblockA{\textit{School of Computer Science and Engineering} \\
\textit{Macau University of Science and Technology}\\
Macau, China \\
liuaansang@gmail.com}
\and
\IEEEauthorblockN{Tao Zhang}
\IEEEauthorblockA{\textit{School of Computer Science and Engineering} \\
\textit{Macau University of Science and Technology}\\
Macau, China \\
tazhang@must.edu.mo}
}

\maketitle

\begin{abstract}
Bug tracking systems store many bug reports, some of which are related to security. Identifying those security bug reports (SBRs) may help us predict some security-related bugs and solve security issues promptly so that the project can avoid threats and attacks. However, in the real world, the ratio of security bug reports is severely low; thus, directly training a prediction model with raw data may result in inaccurate results. Faced with the massive challenge of data imbalance, many researchers in the past have attempted to use text filtering or clustering methods to minimize the proportion of non-security bug reports (NSBRs) or apply oversampling methods to synthesize SBRs to make the dataset as balanced as possible. Nevertheless, there are still two challenges to those methods: 1) They ignore long-distance contextual information. 2) They fail to generate an utterly balanced dataset. To tackle these two challenges, we propose SEDAC, a new SBR identification method that generates similar bug report vectors to solve data imbalance problems and accurately detect security bug reports. Unlike previous studies, it first converts bug reports into individual bug report vectors with distilBERT, which are based on \emph{word2vec}. Then, it trains a generative model through conditional variational auto-encoder (CVAE) to generate similar vectors with security labels, which makes the number of SBRs equal to NSBRs’. Finally, balanced data are used to train a security bug report classifier. To evaluate the effectiveness of our framework, we conduct it on 45,940 bug reports from Chromium and four Apache projects. The experimental results show that SEDAC outperforms all the baselines in \emph{g-measure} with improvements of around 14.24\%-50.10\%.
\end{abstract}

\begin{IEEEkeywords}
security bug report, data imbalance, distilBERT, conditional variational auto-encoder
\end{IEEEkeywords}

\section{Introduction}
The bug tracking system can effectively manage the entire cycle of submission, repair, and closure in the software development process, including reporting bugs, querying bug records, generating reports, etc. These bugs have different classifications and levels. Some bugs can be described as security-related bug reports, which are marked as security bug reports (SBRs), while others are marked as non-security bug reports (NSBRs). Effective use of these bug information can provide developers with security bug predictions and prevent projects from being maliciously attacked in a timely manner.

There is a traditional workflow of SBR identification, which is mainly divided into three stages: 

\emph{\textbf{The first stage}} is text representation for bug reports because the text information in bug reports is difficult to use directly in identification involving deep learning methods. 

\emph{\textbf{The second stage}} is to solve the data imbalance problem between SBRs and NSBRs. In the real world, the ratio of SBRs to NSBRs is seriously out of balance, such as bug report data from Apache Wicket\footnote{\url{https://issues.apache.org/jira/browse/WICKET}}, Ambari\footnote{\url{https://issues.apache.org/jira/browse/AMBARI}}, Camel\footnote{\url{https://issues.apache.org/jira/browse/CAMEL}}, Derby\footnote{\url{https://issues.apache.org/jira/browse/DERBY}}. This severe data imbalance must be addressed if the information from security bug reports is to be used to provide developers with accurate predictions. Mainstream works \cite{petersTextFilteringRanking2019, shuBetterSecurityBug2019, jiangLTRWESNewFramework2020, maCASMSCombiningClustering2022} continue to revolve around the data imbalance problem to do research. 

\emph{\textbf{The third stage}} is to classify the data set obtained after solving the data imbalance problem. The mainstream method is to use basic machine learning algorithms directly because it is the most straightforward, direct, effective, and efficient solution to the binary classification problem. There are also some researchers who are trying to add deep learning methods, such as CNN, Bi-LSTM, or attention mechanisms, in the classification stage.

Several methods \cite{petersTextFilteringRanking2019, shuBetterSecurityBug2019, jiangLTRWESNewFramework2020, maCASMSCombiningClustering2022} were proposed for SBR identification, and all of them have made practical improvements, but there are still two challenges:

\emph{Challenge 1. Ignoring long-distance contextual information:} Previous studies \cite{petersTextFilteringRanking2019, shuBetterSecurityBug2019, jiangLTRWESNewFramework2020, maCASMSCombiningClustering2022} mainly used \emph{tf-idf} and \emph{word2vec} \cite{mikolovDistributedRepresentationsWords2013,mikolovEfficientEstimationWord2013,rongWord2vecParameterLearning2014} to represent bug reports. Both of these algorithms have their advantages and disadvantages. The \emph{tf-idf} algorithm can avoid operations in high-dimensional spaces and obtain representations simply and quickly, but it can easily lead to vector sparsity problems. Although \emph{word2vec} is a very versatile word embedding method, it makes minimal use of the contextual relationship of words.

\emph{Challenge 2. Failed to generate an utterly balanced dataset:} Current approaches \cite{petersTextFilteringRanking2019, shuBetterSecurityBug2019, jiangLTRWESNewFramework2020, maCASMSCombiningClustering2022} to SBR identification primarily fall into two categories from the perspectives of filtering out NSBRs and synthesizing SBRs. The methods to filter NSBRs mainly focus on NSBRs that are the most similar to SBRs, using security-related keywords or text similarity to rank NSBRs and filter them out. The latest research uses \emph{k-means} to cluster NSBRs and group them with SBRs. The synthesis method randomly synthesizes SBRs between every two nearest neighbors of them, using the Synthetic Minority Oversampling Technique (SMOTE) \cite{chawlaSMOTESyntheticMinority2002}. Neither filtering nor synthesis methods generate a balanced data set.

To overcome the challenges, we propose \emph{SEDAC, a \textbf{SE}curity bug report identification framework composed of \textbf{D}istilBERT \textbf{A}nd \textbf{C}onditional variation autoencoder}. To tackle \emph{Challenge 1}, SEDAC first converts bug reports into individual bug report vectors using distilBERT \cite{sanhDistilBERTDistilledVersion2019}, a distilled version of BERT \cite{devlinBertPretrainingDeep2018}, which uses Transformer \cite{vaswaniAttentionAllYou2017} to obtain richer text information and capture longer-distance contextual relationships. Each vector corresponds to a one-hot encoding label. To tackle \emph{Challenge 2}, we introduce conditional variation auto-encoder (CVAE) \cite{sohnLearningStructuredOutput2015}. CVAE is a generative model that consists of an encoder and a decoder. We use the original vectors and corresponding labels to train and obtain the well-trained CVAE model. Then, the decoder from CVAE is separated and fed with random vectors from Gaussian distribution and specified condition vectors to generate new bug report vectors. After the balance data is obtained, it can be sent to train an SBR classifier using \emph{Logistic Regression (LR)}, which is proven to perform better than other machine learning algorithms (See Section \ref{5.3.2}).

To evaluate the performance of SEDAC, we conduct experiments on five open-source projects: four from Apache projects, each containing 1,000 bug reports, and one from Chromium projects, containing 41,940 bug reports. K-fold cross-validation method \cite{rodriguezSensitivityAnalysisKfold2009, wongReliableAccuracyEstimates2019} is adopted to avoid the impact of random seed on classifier training. The experimental results on benchmark datasets show that SEDAC outperforms the state-of-the-art method on average by 14.24\%-50.10\% in \emph{g-measure} and 6.99\%-45.04\% in \emph{pd}, respectively, with maximum \emph{pf} of no more than 1\%.

Overall, there are three contributions of our paper:

\begin{itemize}
\item \textbf{Significance.} We provide a new perspective to solve the data imbalance problem in security defect report identification, synthesizing a completely balanced data set, which benefits the identification work.
\item \textbf{Approach.} We introduce SEDAC, a novel SBR detection framework that can automatically extract diverse bug reports and generate security bug reports to improve the performance of SBR identification.
\item \textbf{Experiments.} We conduct experiments on five real-world projects, and empirical results prove that SEDAC is more efficient for identifying security bug reports. We have made our work publicly accessible at \url{https://github.com/YanshengLiao/SEDAC}.
\end{itemize}

The remainder of the paper is organized as follows. Section II introduces the related work of SBR identification. Section III presents the background knowledge essential for understanding our approach. Section IV illustrates our framework in detail. Section V evaluates the effectiveness of our framework. Section VI discusses the reasons for improvement and the limitations of our method. Finally, Section 7 concludes the paper.

\section{Related Work}

Bug tracking systems collect bug reports from various sources, such as development teams, testing teams, and end users. Reporters usually have to determine whether the submitted bug reports involve security issues and mark these security-related bug reports as security bug reports (SBRs). However, in the bug reporting process, bug reporters often mislabel SBRs as NSBRs partly due to a lack of security domain knowledge. These mislabeled security bug reports can easily lead to delayed identification and remediation of vulnerabilities. Gegick et al. \cite{gegickIdentifyingSecurityBug2010a} first raised this problem and developed a text mining method to identify SBRs manually incorrectly labeled as NSBRs.

However, after solving the problem of mislabeling, it is still difficult for engineers to use past bug report information to accurately predict future bugs because, in reality, the proportion of SBRs in bug reports is deficient. As shown in Table \ref{tbl2}, among the 45,940 bug reports from Chromium and four projects in Apache, the proportion of SBRs does not exceed 10\%.

To solve the data imbalance problem, Peters et al. \cite{petersTextFilteringRanking2019} first proposed FARSEC, a framework that filters out NSBRs through security-related words and uses the idea of blending to build a prediction model. It first filters out the top 100 security-related keywords from SBRs through \emph{tf-idf} and then uses them to score and rank NSBRs. NSBRs with high scores are filtered out because it means they are very similar to SBRs. Afterward, Goseva et al. \cite{goseva-popstojanovaIdentificationSecurityRelated2018} proposed a supervised approach and an unsupervised approach to identify security bug reports. They created feature vectors for bug reports using \emph{bf}, \emph{tf}, and \emph{tf-idf}. Then, they combined the vectors with supervised and unsupervised classification methods. Shu et al. \cite{shuBetterSecurityBug2019} improved FARSEC by combining SMOTE with the filters of it. Instead of filtering out NSBRs, SMOTE calculates the \emph{k} nearest neighbors for each minority class sample, that is, SBRs, and oversamples to make more of them. Additionally, they optimize the hyperparameters of classification learners. In subsequent work, researchers continue to improve methods for detecting SBRs. Jiang et al. \cite{jiangLTRWESNewFramework2020} proposed LTRWES, a method for filtering NSBRs based on text similarity using $BM25F_{ext}$ \cite{sunMoreAccurateRetrieval2011}. The method of LTRWES is similar to FARSEC, which changes the focus of filtering from security-related words to text similarity. Ma et al. \cite{maCASMSCombiningClustering2022} proposed CASMS, a method for filtering NSBRs based on \emph{k-means}. CASMS divides NSBRs into groups through \emph{k-means}, combining different groups with SBRs as training data. In addition, it adds a mixed model CNN-BLSTM \cite{wuAutomaticAudioChord2018, liuProtDetCCHProteinRemote2018} and Attention mechanism \cite{bahdanauNeuralMachineTranslation2014} to enhance the text information of bug reports.

\begin{table*}
\renewcommand{\arraystretch}{1.3}
\caption{Summary of related works and SEDAC.}
\label{tbl1}
\resizebox{1.0\linewidth}{!}{
\begin{tabular}{cccc}
\toprule
Framework & \makecell[c]{Stage 1\\Text Representation} & \makecell[c]{Stage 2\\Measures for Data Imbalance\\(\textbf{F}ilter/\textbf{S}ythesize)} & \makecell[c]{Stage 3\\Classification}    \\ 
\midrule
FARSEC & $tf$-$idf$ & Security-related keywords (F) & Machine learning  \\
SMOTUNED & $tf$-$idf$ & SMOTE (S) & Parameter optimization for machine learning  \\
LTRWES & $tf$-$idf$ & Text similarity using $BM25F_{ext}$ (F) & Machine learning  \\
CASMS & $word2vec$ \& $tf$-$idf$ & K-Means (F) & CNN-BLSTM \& Attention   \\
\cdashline{1-4}[1pt/1pt]
SEDAC & distilBERT & CVAE model (S) & Machine learning  \\
\bottomrule
\end{tabular}
}
\end{table*}

As shown in Table \ref{tbl1}, there is a summary of related works mentioned above and the proposed framework SEDAC. The supervised approach and the unsupervised approach proposed by Goseva et al. \cite{goseva-popstojanovaIdentificationSecurityRelated2018} are not included because their experiments were conducted on three datasets from NASA instead of the five projects used by the other four frameworks. This table summarizes that: 1) Previous studies revolve around \emph{tf-idf} in the stage of text representation. 2) The primary measure to solve the data imbalance problem is to filter NSBRs using various methods. 3) Machine learning algorithms are mainly used for classification, and some researchers attempt to make innovations.

The SEDAC that we propose differs from the approaches mentioned above; distilBERT is adopted in the stage of text representation, and the CVAE model is used to synthesize SBRs.

\section{Background}

\subsection{Bug Reports}

\begin{figure}
	\centering
		\includegraphics[width=1.0\linewidth]{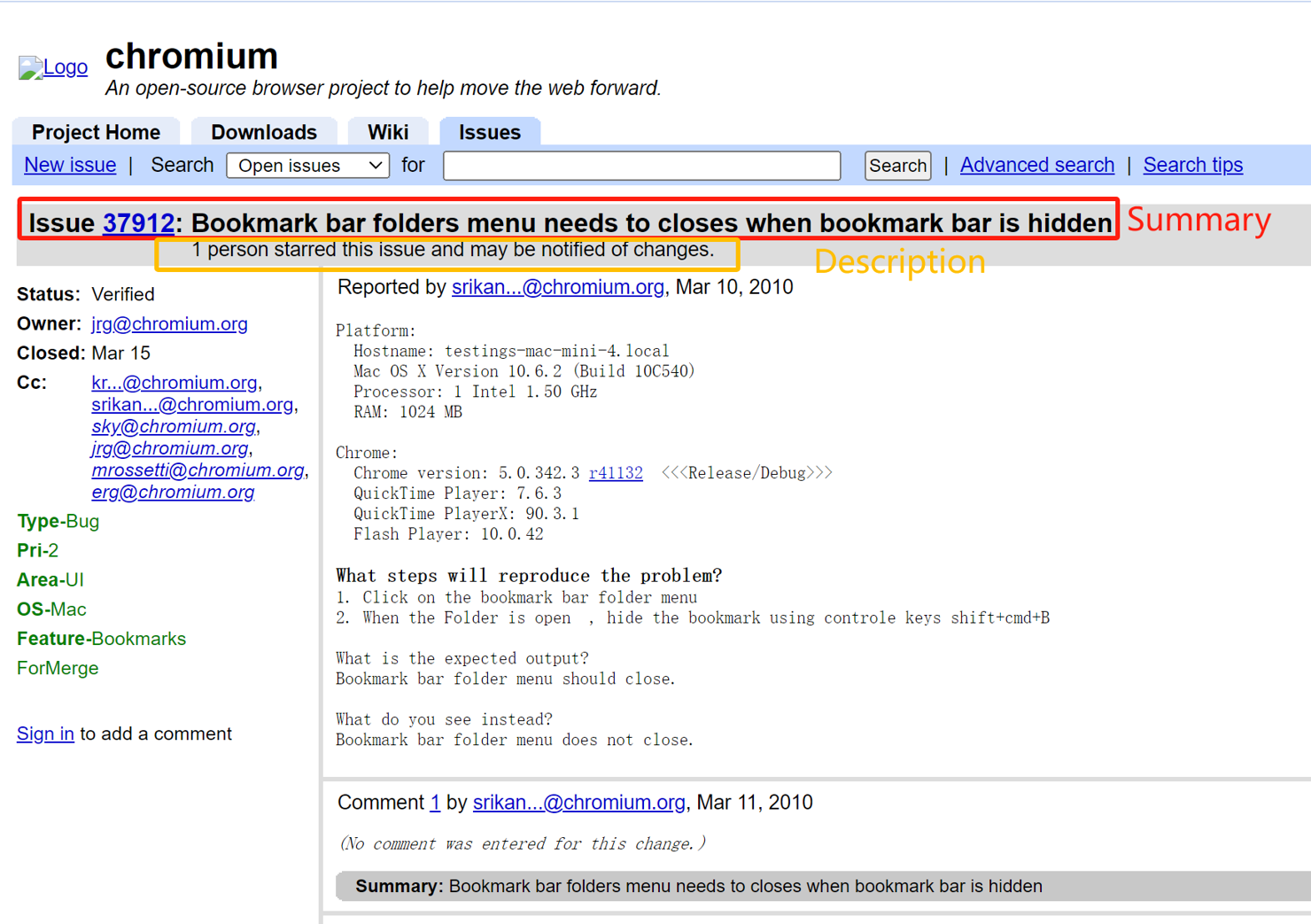}
	  \caption{An example of a bug report from the Chromium project.}\label{fig1}
\end{figure}

A bug report is a record written by software developers after fixing the bugs during software maintenance \cite{zhangLiteratureReviewResearch2016}, which is generally stored in a bug-tracking system. An effective bug report should include title, environment (OS, software version, etc.), steps to reproduce a bug, expected results, actual results, etc. A good bug report can cover all essential information about the bug, help with detailed bug analysis, help developers find the correct direction and method for program debugging, and help avoid the same bugs in future versions. In the research field of software engineering, there are many research directions related to bug reports, such as bug localization \cite{sahaImprovingBugLocalization2013, zhouWhereShouldBugs2012}, duplicate bug report detection \cite{nguyenDuplicateBugReport2012, budhirajaDWENDeepWord2018}, bug report summarization \cite{rastkarAutomaticSummarizationBug2014, maniAusumApproachUnsupervised2012}, etc. Figure \ref{fig1} shows an example of a bug report from the Chromium project. The framed summary and description are the text information used in our method. Researchers with specific knowledge in software engineering can determine the type of a bug through the summary and description, such as usability, security, functionality, logic, etc. Our approach studies the identification of security bug reports.

\subsection{Bidirectional Encoder Representations from Transformers and Its Variants}\label{3.2}

Bidirectional Encoder Representation from Transformers (BERT) \cite{devlinBertPretrainingDeep2018} is a language representation model designed to pre-train deep bidirectional representations from the unlabeled text by joint conditioning on both left and right context in all layers. It consists of 12 layers of transformer blocks with a hidden size 768. Unlike traditional left-to-right or right-to-left language models, BERT can do both simultaneously. It is trained with two unsupervised tasks: Masked LM (MLM) and Next Sentence Prediction (NSP). For the MLM task, BERT randomly masks some percentage of the input tokens and then predicts those masked tokens. NSP task is to establish a predictive relationship between the previous sentence and the following one. As a pre-trained language model, it can be connected to various downstream tasks after BERT, such as text classification, question answering, etc., and then fine-tuned. 

After BERT came out, researchers tried to make different improvements to BERT, resulting in various variants of BERT. There are some common variants, such as RoBERTa \cite{liuRobertaRobustlyOptimized2019}, ALBERT \cite{lanAlbertLiteBert2019}, distilBERT \cite{sanhDistilBERTDistilledVersion2019}, distilRoBERTa, ELECTRA \cite{clarkElectraPretrainingText2020}, etc. RoBERTa \cite{liuRobertaRobustlyOptimized2019} uses a more extensive byte-level BPE dictionary at the text encoding stage, which has solved the out-of-vocabulary (OOV) problem \cite{sennrichNeuralMachineTranslation2015} and proposes a dynamic masking strategy. ALBERT \cite{lanAlbertLiteBert2019} is a lite BERT model that proposes two methods to decrease the parameters to alleviate the memory consumption problem. DistilBERT \cite{sanhDistilBERTDistilledVersion2019} uses knowledge distillation to accelerate the pre-trained model's reasoning and reduce the model's size based on ensuring the effect of the model. DistilRoBERTa is also a distilled version of RoBERTa, following the same training procedure as distilBERT \cite{sanhDistilBERTDistilledVersion2019}. ELECTRA \cite{clarkElectraPretrainingText2020} introduces a generator and discriminator similar to the confrontation network (GAN) to update model parameters more efficiently. Many studies \cite{zhangEmotionalClassificationMethod2022, liaoImprovedAspectcategorySentiment2021} have proved that all the variants mentioned have outstanding performance in text representation.

In our approach, distilBERT is applied to convert bug reports into vector representations. It performs slightly better in SEDAC than other BERT variants.

\subsection{Conditional Variational Auto-Encoder}
\begin{figure}
	\centering
		\includegraphics[width=0.8\linewidth]{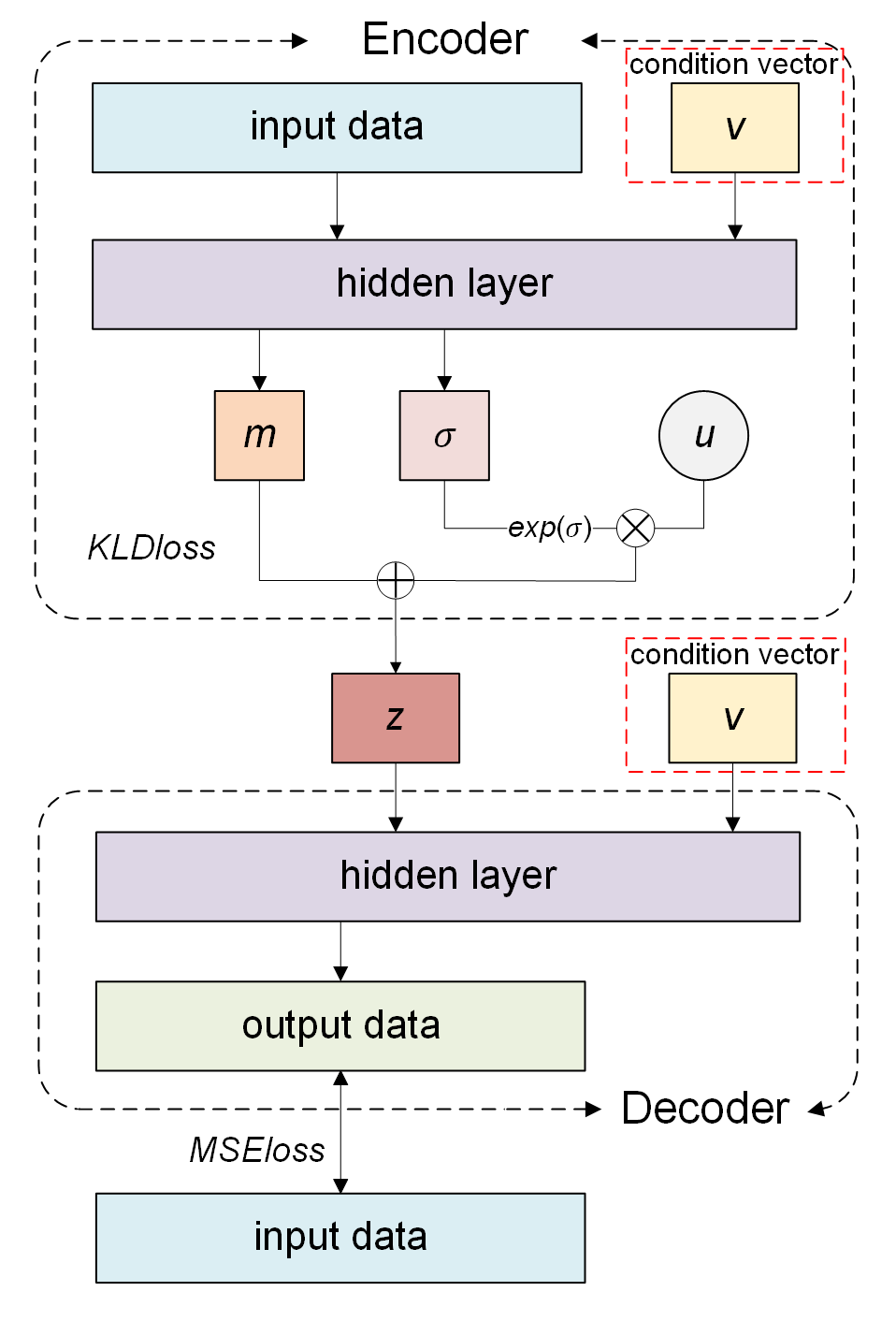}
	  \caption{The difference between VAE and CVAE.}\label{fig2}
\end{figure}

The core of SEDAC is to increase the number of SBRs through the data augmentation method to make the data set completely balanced. With the development of deep generative models, we consider conditional variational autoencoders (CVAE) more suitable for our framework.

Before the emergence of CVAE, there were two major generative models, namely Generative Adversarial Nets (GAN) \cite{goodfellowGenerativeAdversarialNets2014} and variational autoencoder (VAE) \cite{kingmaAutoencodingVariationalBayes2013}. GAN, consisting of discriminator and generator nets, is a popular neural net. It is based on game theory and aims to find the Nash Equilibrium \cite{krepsNashEquilibrium1989} in the discriminator and generator networks. Many combinations with GAN, like DCGAN \cite{radfordUnsupervisedRepresentationLearning2015}), Sequence-GAN \cite{yuSeqganSequenceGenerative2017}, LSTM-GAN \cite{zhuNovelLSTMGANAlgorithm2019}, etc., have shown outstanding results, which also means that the discriminator and generator play a very influential role in the adversarial generative network. The generating and training modes of VAE are very different from GAN. VAE is based on Bayesian inference \cite{griffithsBayesianInference2012, kingmaAutoencodingVariationalBayes2013}, modeling the potential distribution of data and then sampling new data from the potential distribution. Latent variables can be likened to a summary description of the input data, which summarizes what kind of data we want. To a certain extent, VAE is an expression model that specifies a standard and template for the data that must be generated through latent variables so that the model does not unthinkingly generate data. As shown in Figure \ref{fig2} (contents in the red dotted box are excluded), the VAE model consists of an encoder and a decoder. The encoder samples the input data through the hidden layer and obtains the Gaussian distribution of the sample data that obeys the corresponding mean and variance.

However, whether it is GAN or VAE, their generation process is random and uncontrollable. Thus, conditional variational automatic encoding (CVAE) came into being. As illustrated in Figure \ref{fig2}, CVAE has one more condition vector than VAE, which refers to the label of the input data and is a form of one-hot encoding. The condition vector and the input data are combined to train the encoder and decoder. The training goal of the model is that the decoded data should be similar to the input data. Due to the existence of a condition vector, the decoder can be separated from the entire CVAE model and input random vectors and given labels to generate the data we want.

\section{Approach}
\subsection{Overview}

\begin{figure*}
	\centering
		\includegraphics[width=1.0\textwidth]{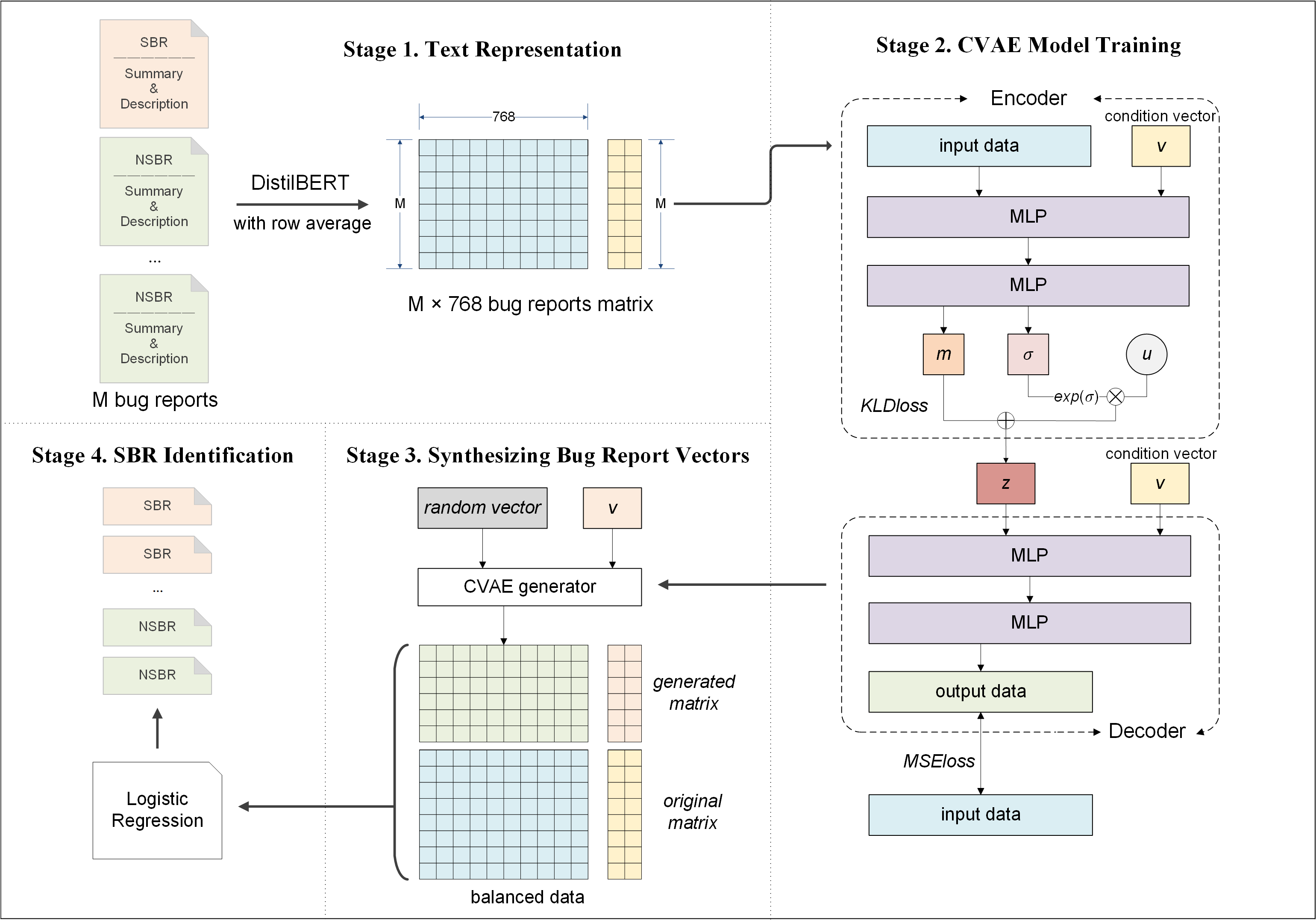}
	  \caption{Overview of SEDAC.}\label{fig3}
\end{figure*}

The framework of SEDAC is shown in Figure \ref{fig3}, divided into four stages: text representation, model training, data synthesis, and SBR identification. In the first stage, SEDAC uses distilBERT to convert the text in bug reports into bug report vectors. In the second stage, SEDAC obtains the decoder for data synthesis by training the CVAE model. In the third stage, SEDAC feeds random vectors and specific labels as input data into the obtained decoder for data synthesis. The synthesized data in this stage is 1:1 balanced. In the fourth stage, the fully balanced dataset is used as input data to train the SBR identification model.

\subsection{Text Representation}

Bug reports contain rich information, such as summary, description, resolution, priority, status, etc. The textual information in the summary and description best summarizes whether a bug report is security-related. Pre-trained distilBERT can be used to encode the text. Therefore, we merge the summary and description contents and perform word embedding on them through distilBERT. The max embedding length of distilBERT is 512; thus, we uniformly adopt truncation measures for a text that exceeds 512 embedding lengths. As mentioned before (See Section \ref{3.2}), distilBERT contains 768 hidden units, which means that each obtained embedding word is a tensor with the size of $1\times768$. Hence, the embedding representation of a sentence containing N words is a matrix of size $N\times768$. However, whether the matrix sizes are different due to different sentence lengths or the matrix is too heavy, it is not conducive to subsequent model training. To simplify the embedding representation, we perform row averaging on each sentence representation so each bug report can be represented as a vector with a size of $1\times768$. For $M$ bug reports, they are converted into a matrix with a size of $M\times768$. Specifically, the labels of SBRs (i.e., the value of 1) and NSBRs (i.e., the value of 0) result in one-hot vectors [0, 1] and [1, 0], respectively.

In short, each bug report is turned into an independent text vector after processing, which is used as input data for model training, with the label as a condition vector in one-hot encoding form.

\subsection{CVAE Model Training}

As shown in Figure 3, the second part is training the CVAE model using an encoder-decoder structure. In the encoder, the bug report vectors and the corresponding one-hot condition vectors serve as input to our model. Followed by two layers of MLPs, they form a fully connected network, matching each input data with a Gaussian distribution that obeys the corresponding mean $m$ and standard deviation $\sigma$. The $u$ is a noise vector randomly sampled from the Gaussian distribution. The encoder generates the latent variables z by Equation 1 \cite{doerschTutorialVariationalAutoencoders2016}.
\begin{equation}z=m+e^\sigma\times u\end{equation}

To prevent the value of $e^\sigma$ from being infinitely close to 0, the $KLDloss$ function is set in the sampling process. It can be calculated by Equation 2 \cite{doerschTutorialVariationalAutoencoders2016}.
\begin{equation}KLDloss=-\sum\frac{1+\sigma-m^2-e^t}{2}\end{equation}

In the decoder, the network of two-layer MLPs is learned to generate the bug report vectors from the sampled distribution. The mean square error (MSE) ensures the generated quality of the reconstructed bug report vectors. It can be calculated by Equation 3 \cite{doerschTutorialVariationalAutoencoders2016}. $X(i)$ refers to the initial input data at the very beginning of CVAE model training. $X_{reconstruct}(i)$ refers to the output data in the decoder.
\begin{equation}MSEloss=\frac{\sum_{j=1}^{768}{(X(i)-X_{reconstruct}(i))}^2}{768}\end{equation}

The training process aims to ensure that the reconstructed bug report vectors are as consistent as possible with the input data by minimizing $KLDloss$ and $MSEloss$.

\subsection{Synthesizing Bug Report Vectors}

The decoder plays a vital role in the data synthesis stage. It is separated from the well-trained CVAE model as a generator and synthesizes bug report vectors of specified labels. Specifically, we can generate synthesized bug report vectors by feeding the decoder with a random vector from the Gaussian distribution and a specified condition vector. The condition vector should be [0,1] because the data we need are SBRs. Note that at this stage, the amount of generated data is calculated and is the difference between NSBRs and SBRs. It also means that the generated vectors combined with the original vectors become a fully balanced data set with SBRs: NSBRs = 1:1, which is beneficial to subsequent classification tasks.

\subsection{SBR Identification}

The generated data output from the synthesis stage is combined with the original data as the input data for SBR identification. Our identification model of choice is logistic regression, which performs better than other machine learning algorithms (See Section \ref{5.3.2}).

\section{Experiment}

Our experiment was conducted on an AMD EPYC 7742 64-core Processor with 96G RAM. 

\subsection{Datasets}

\begin{table*}
\caption{Details of collected projects.}
\label{tbl2}
\resizebox{1.0\linewidth}{!}{
\begin{tabular}{llcccc}
\toprule
Project & Domain & Time Period & BRs & SBRs & SBRs(\%)    \\
\midrule
Chromium & Web browser called Chrome. & 08/30/2008-06/11/2010 & 41,940 & 192 & 0.5    \\
Wicket & Component-based web application framework for Java programming. & 10/20/2006-11/09/2014 & 1,000 & 10 & 1.0  \\
Ambari & Hadoop management web UI backed by its RESTful APIs.& 09/26/2011-08/08/2014 & 1,000 & 29 & 3.0  \\
Camel & A rule-based routing and mediation engine. & 07/08/2007-09/18/2013 & 1,000 & 32 & 3.0 \\
Derby & A relational database management system. & 09/28/2004-09/17/2014 & 1,000 & 88 & 9.0  \\
\bottomrule
\end{tabular}
}
\end{table*}

We evaluate SEDAC on five real-world projects with historical bug reports that were labeled SBRs and NSBRs: four from Ohira et al. \cite{ohiraDatasetHighImpact2015}, and a subset of bug reports from the Chromium project, compiled by Peters et al. \cite{petersTextFilteringRanking2019}. Four open-source Apache projects are applied in different domains and are tracked by the bug tracking system JIRA\footnote{\url{https://www.atlassian.com/software/jira/}}. Ohira et al. \cite{ohiraDatasetHighImpact2015} supply them in Microsoft Excel format, organized into comma-separated value (CSV) files by Peters et al. The Chromium dataset comes from the 2011 mining challenge of the Mining Software Repositories conference\footnote{\url{http://2011.msrconf.org/msr-challenge.html}} \cite{petersTextFilteringRanking2019}. Security bugs are labeled as \emph{Bug-Security} while other types of bugs are treated as non-security bug reports. We use the same datasets as baselines for a fair comparison. Table \ref{tbl2} shows each project's domain and other details. We have not updated the dataset to the latest date to provide a fair comparison with the state-of-the-art work.

\subsection{Evaluation Metrics}

\begin{table}
\centering
\renewcommand{\arraystretch}{1.4}
\caption{Evaluation metrics: confusion matrix, definitions of \emph{pd}, \emph{pf} and \emph{g-measure}.}
\label{tbl3}
\begin{tabular}{c|c|c|c|}
\multicolumn{2}{c}{} & \multicolumn{2}{c}{Actual} \\
\cline{3-4}
\multicolumn{2}{c|}{} & SBRs & NSBRs \\
\cline{2-4}
\multirow{2}*{Predict} & SBRs & \textit{TP} & \textit{FN} \\ 
\cline{2-4}
 & NSBRs & \textit{FP} & \textit{TN} \\
\cline{2-4}
 & pd & \multicolumn{2}{c|}{$\frac{TP}{TP+FN}$} \\
 & pf & \multicolumn{2}{c|}{$\frac{FP}{FP+TN}$} \\
 & g-measure & \multicolumn{2}{c|}{$\frac{2\times pd \times(100-pf)}{pd+(100-pf)}$} \\
\cline{2-4}
\end{tabular}
\end{table}

To evaluate the performance of the classification models, we use the measures shown in Table \ref{tbl3}. The confusion matrix is used, where \emph{TP}, \emph{FP}, \emph{FN} and \emph{TN} are true positive, true negative, false positive, and false negative respectively:
\begin{itemize}
\item \emph{TP:} The number of SBRs that are correctly predicted.
\item \emph{FP:} The number of NSBRs that are mistakenly predicted as SBRs.
\item \emph{FN:} The number of SBRs that are mistakenly predicted as NSBRs.
\item \emph{TN:} The number of NSBRs that are correctly predicted.
\end{itemize}

We mainly focus on probability of detection \emph{pd}, probability of false alarm \emph{pf}, and \emph{g-measure} \cite{jiangTechniquesEvaluatingFault2008a}:

\begin{itemize}
\item \emph{pd:} the fractions of SBRS that are correctly predicted.
\item \emph{pf:} the ratio of the NSBRs that are incorrectly predicted as SBRs.
\item \emph{g-measure:} the harmonic mean of $pd$ and $(100-pf)$. The $100-pf$ is known as a concept named specificity, which is defined as the fractions of not predicting NSBRs as SBRs. It is proved that the $g$-$measure$ is more suitable for unbalanced data \cite{petersTextFilteringRanking2019,jiangTechniquesEvaluatingFault2008a}.
\end{itemize}

\subsection{Research Questions and Results}
\subsubsection{RQ1.How does SEDAC perform in SBR identification compared with the state-of-the-art methods?}

\paragraph{Baselines}
In order to evaluate the effect of SEDAC, the following frameworks are selected as baselines for comparison with SEDAC:

\begin{itemize}
\item \textbf{FARSEC:} FARSEC \cite{petersTextFilteringRanking2019} uses \emph{tf-idf} to identify security-related keywords and filter out NSBRs based on them. It finally uses the idea of blending to rank bug reports.
\item \textbf{SMOTUNED:} SMOTUNED \cite{shuBetterSecurityBug2019} is an improvement of FARSEC; it combines SMOTE with FARSEC and optimizes the hyper-parameters of machine learning algorithms. 
\item \textbf{LTRWES:} LTRWES \cite{jiangLTRWESNewFramework2020} is also an improvement of FARSEC. It calculates text similarity based on $BM25F_{ext}$ to filter out NSBRs that are most similar to SBRs, replacing the method in FARSEC that uses security-related keywords to filter NSBRs.
\item \textbf{CASMS:} CASMS \cite{maCASMSCombiningClustering2022} clusters NSBRs into different classes through \emph{k-means}, and then combines different NSBR groups and SBRs into a smaller but relatively balanced data set for SBR identification. During the process, it also combines the CNN-BLSTM model and attention mechanism.
\end{itemize}

In the process of reproducing FARSEC, we could not reproduce it smoothly due to the problems with the \emph{Clojure} language version. Similarly, when we reproduced LTRWES, we could not find some packages used in their code, thus we could not reproduce it, either. For SMOTUNED and CASMS, the code is not given in the article. Therefore, we uniformly select the results with the best g-measure performance among their results to compare with the results of our method. Since the data sets used by our method are consistent with those used by baselines, such comparison is reliable.

\paragraph{Results}

\begin{table*}
\centering
\caption{Comparison of FARSEC, SMOTUNED, LTRWES, CASMS, and SEDAC in \emph{g-measure(\%)}, \emph{pd(\%)}, and \emph{pf(\%)}.}
\label{tbl4}
\resizebox{0.7\linewidth}{!}{
\begin{tabular}{cccccccc}
\toprule
Project & Metric & FARSEC & SMOTUNED & LTRWES & CASMS & SEDAC & Improvement \\
\midrule
\multirow{3}*{Chromium} & g-measure & 65.40 & 80.53 & 86.67 & 74.00 & \textbf{99.77} & \textcolor{red!95}{$\uparrow$15.11\%} \\
 & pd & 49.60 & 74.80 & 93.04 & 73.05 & \textbf{99.54} & \textcolor{red!95}{$\uparrow$6.99\%}  \\
 & pf & 3.80 & 12.80 & 18.88 & 24.41 & \textbf{0.00} & -  \\
\midrule
\multirow{3}*{Wicket} & g-measure & 65.00 & 75.85 & 70.93 & 64.60 & \textbf{99.49} & \textcolor{red!95}{$\uparrow$31.17\%}  \\
 & pd & 66.70 & 66.70 & 83.33 & 73.33 & \textbf{98.99} & \textcolor{red!95}{$\uparrow$18.79\%}  \\
 & pf & 36.60 & 12.10 & 38.26 & 40.85 & \textbf{0.00} & -  \\
\midrule
\multirow{3}*{Ambari} & g-measure & 71.90 & 71.64 & 86.16 & 83.12 & \textbf{98.43} & \textcolor{red!95}{$\uparrow$14.24\%}  \\
 & pd & 57.10 & 57.10 & 85.71 & 82.85 & \textbf{97.12} & \textcolor{red!95}{$\uparrow$13.31\%}  \\
 & pf & 3.00 & 3.90 & 13.39 & 16.10 & \textbf{0.21} & -  \\
\midrule
\multirow{3}*{Camel} & g-measure & 53.80 & 64.43 & 65.47 & 62.99 & \textbf{98.27} & \textcolor{red!95}{$\uparrow$50.10\%}  \\
 & pd & 50.00 & 55.60 & 66.67 & 64.45 & \textbf{96.70} & \textcolor{red!95}{$\uparrow$45.04\%}  \\
 & pf & 41.80 & 23.40 & 35.68 & 37.05 & \textbf{0.10} & -  \\
\midrule
\multirow{3}*{Derby} & g-measure & 61.70 & 69.60 & 75.92 & 74.72 & \textbf{95.80} & \textcolor{red!95}{$\uparrow$26.19\%}  \\
 & pd & 47.60 & 57.10 & 69.05 & 70.00 & \textbf{92.65} & \textcolor{red!95}{$\uparrow$32.36\%}  \\
 & pf & 12.40 & 10.90 & 15.69 & 19.44 & \textbf{0.77} & -  \\
\midrule
\multirow{3}*{\textbf{Average}} & g-measure & 63.56 & 72.41 & 77.03 & 71.89 & \textbf{98.35} & \textcolor{red!95}{$\uparrow$27.68\%}  \\
 & pd & 54.20 & 62.26 & 79.56 & 72.74 & \textbf{97.00} & \textcolor{red!95}{$\uparrow$21.92\%}  \\
 & pf & 19.52 & 12.62 & 24.38 & 27.57 & \textbf{0.22} & -  \\	
\bottomrule
\end{tabular}
}
\end{table*}

As can be seen from Table \ref{tbl4}, for the \\Chromium project, the best baseline result comes from LTRWES, whose \emph{g-measure} is 86.67\% and \emph{pd} is 93.04\%. The results of SEDAC are that \emph{g-measure} is 99.77\% and \emph{pd} is 99.54\%, and the corresponding improvements are 15.11\% and 6.99\% respectively. For Wicket, the highest \emph{g-measure} comes from SMOTUNED, which is 75.85\%, while SEDAC’s \emph{g-measure} is 99.49\%, with an improvement of 31.17\%. The highest \emph{pd} comes from LTRWES, 83.33\%, while the \emph{pd} of SEDAC is 98.99\%, with an improvement of 18.79\%. For Ambari, the highest \emph{g-measure} and \emph{pd} are both from LTRWES, which are 86.16\% and 85.71\% respectively, while the results of SEDAC are 98.43\% and 97.12\%, with improvements of 14.24\% and 13.31\% respectively. For Camel, all baseline results perform poorly, and only the results of LTRWES are slightly better than others, which are 65.47\% in \emph{g-measure} and 66.67\% in \emph{pd} respectively. SEDAC's performance in Camel is still very good, with 98.27\% in \emph{g-measure} and 96.70\% in \emph{pd}, accompanied by improvements of 50.10\% and 45.04\% respectively. For Derby, the highest \emph{g-measure} and highest \emph{pd} are from LTRWES and CASMS, which are 75.92\% and 70.00\% respectively. The results of SEDAC are 95.80\% and 92.6\%, with improvements of 26.19\% and 32.36\%. For all projects, the \emph{pf} results of SEDAC are as low as 0\%, as high as 0.77\%, or even no more than 1\%. On average, SEDAC's results improve over the best baselines by 27.68\% in \emph{g-measure} and 21.92\% in \emph{pd}. In summary, SEDAC outperforms all baselines.

\subsubsection{RQ2.How effective is the combination of distilBERT(uncased) and Logistic Regression compared to other language models and machine learning algorithms?}\label{5.3.2}
\paragraph{Text representation}
In the stage of text representation stage, distilBERT (uncased) is used to capture the contextual information of bug reports and convert them into vectors. It is a distilled and case-insensitive version of BERT. We select some other common BERT variants to compare with distilBERT (uncased) as follows:
\begin{itemize}
\item \textbf{BERT (cased):} BERT \cite{devlinBertPretrainingDeep2018} is an unsupervised pre-trained language representation model that stands for Bidirectional Encoder Representations from Transformers. It totalizes 110M parameters and it is case-sensitive.
\item \textbf{BERT (uncased):} An uncased version of BERT. It is case-insensitive.
\item \textbf{RoBERTa:} RoBERTa \cite{liuRobertaRobustlyOptimized2019} has the same architecture as BERT but uses a byte-level BPE as a tokenizer (same as GPT-2) and uses a different pretraining scheme, including dynamic masking, training with larger batches, etc. It is case-sensitive.
\item \textbf{DistilRoBERTa:} DistilRoBERTa is a distilled version of RoBERTa, which follows the same training procedure as distilBERT \cite{sanhDistilBERTDistilledVersion2019}. The model has six layers, totaling 82M parameters. On average, it is twice as fast as RoBERTa-base. It is case-sensitive.
\item \textbf{ALBERT:} ALBERT \cite{lanAlbertLiteBert2019} presents parameter-reduction techniques to lower memory consumption and increase the training speed of BERT. There are two versions of the ALBERT base model, and version 2 is selected because it has better results in nearly all downstream tasks due to different dropout rates, additional training data, and more extended training. It is case-insensitive.
\item \textbf{DistilBERT (cased):} DistilBERT \cite{sanhDistilBERTDistilledVersion2019} is a distilled version of BERT, with 40\% less parameters and 60\% faster than BERT, and preserves over 95\% performance of BERT. It is case-sensitive.
\item \textbf{ELECTRA:} ELECTRA \cite{clarkElectraPretrainingText2020} adjusts the pretraining framework by training two transformer models: the generator and the discriminator. It is case-insensitive.
\end{itemize}

All models are used in their base version due to lower costs with fewer parameters.

\paragraph{Machine learning algorithms}
In the stage of SBR identification, \emph{Logistic Regression} is used to classify the balanced data. It is a traditional statistics technique that models the probability of a discrete outcome. Salem et al. \cite{salemPredictionSoftwareFailures2004} have proved their usefulness for the prediction of software failures. Some common and classic machine learning algorithms are selected to compare with it as follows: 
\begin{itemize}
\item \textit{Support vector machine (SVM)} is a powerful supervised method for classification and regression tasks. It worked well when used by Xing et al. \cite{feixingNovelMethodEarly2005} for software quality prediction, and Ye et al. \cite{yeDDoSAttackDetection2018} for attack detection.
\item \textit{Naive Bayes (NB)} is a classic supervised algorithm for classification tasks. Catal et al. \cite{catalPracticalDevelopmentEclipsebased2011} applied it to software fault prediction, and Ji et al. \cite{jiNewWeightedNaive2019} made use of it for software defect detection and it showed great performance. Considering that our data is continuous, we used Gaussian NB instead of other NB methods.
\item \textit{Random Forest (RF)} is a combination of multiple decision trees. Li et al. \cite{liLargeScaleMaliciousSoftware2021} adopted it on malicious software classification, and it outperformed several state-of-the-art classifiers.
\item \textit{K-Nearest Neighbors (KNN)} is a non-parametric learning algorithm that uses proximity to group data and makes predictions or classifications. Liu et al. \cite{liuNoisyDataElimination2012} proposed a classification algorithm for noisy data elimination based on it, and it performed well.
\item \textit{Multilayer Perceptron (MLP)} is a feed forward artificial neural network. Nassif et al. \cite{nassifEarlySoftwareEstimation2013} developed an MLP neural network model to predict software effort, showing its effectiveness.
\end{itemize}

In the experiment, we use the default parameters of the model. The main reasons are: 1) With the use of the default parameters, the model results have performed well enough (\emph{g-measure} exceeds 95\%). 2) The adjusted parameters for the model that trains on the specified data set are only applicable to that data set. Thus, the parameter adjustment is of little significance.

When dividing the training set and the test set in the early stage of training, different random seeds define different divisions of data sets, resulting in large fluctuations in the training results. In order to reduce the impact of random seeds, we adopt the k-fold cross-validation method. The values of \emph{k} are generally 3, 5, and 10. Considering that for Chromium, 3 is too small, and for the other four projects, 10 may be too large, thus we selected 5 as our \emph{k} value. Specifically, a data set is divided into 5 parts, and 5 training times are performed. Each data set is used as a test set in turn, and the remaining 4 parts are used as training sets. The average of the five results obtained is the final result of the model.

\paragraph{Results}
\begin{table*} 
\centering
\caption{Comparison of different combinations in \emph{g-measure(\%)}, \emph{pd(\%)}, and \emph{pf(\%)}.}  
\label{tbl5}
\resizebox{1.0\linewidth}{!}{
\begin{tabular}{cccccccccccccccccccc}
\Xhline{1pt}
\multirow{2}*{Model} &\multirow{2}*{Project} & \multicolumn{3}{c}{SVM} &  \multicolumn{3}{c}{NB} &  \multicolumn{3}{c}{LR} &  \multicolumn{3}{c}{RF} &  \multicolumn{3}{c}{KNN} &  \multicolumn{3}{c}{MLP} \\ 
\Xcline{3-20}{0.5pt}
& & g-measure & pd & pf & g-measure & pd & pf & g-measure & pd & pf & g-measure & pd & pf & g-measure & pd & pf & g-measure & pd & pf \\
\Xhline{0.7pt} 
\multirow{6}*{\makecell[c]{BERT\\(cased)}} & Chromium & 99.77 & 99.54 & 0.00 & 99.77 & 99.54 & 0.00 & 99.77 & 99.54 & 0.00 & 99.77 & 99.54 & 0.00 & 99.77 & 99.54 & 0.00 & 99.74 & 99.63 & 0.15  \\		
& Wicket & 99.49 & 98.99 & 0.00 & 99.49 & 98.99 & 0.00 & 99.49 & 98.99 & 0.00 & 99.49 & 98.99 & 0.00 & 99.49 & 98.99 & 0.00 & 99.44 & 98.99 & 0.10  \\		
& Ambari & 98.48 & 97.01 & 0.00 & 98.48 & 97.01 & 0.00 & 98.49 & 97.22 & 0.21 & 98.43 & 97.01 & 0.10 & 98.54 & 97.12 & 0.00 & 98.23 & 97.22 & 0.72  \\        
& Camel & 98.32 & 96.70 & 0.00 & 98.32 & 96.70 & 0.00 & 98.27 & 96.70 & 0.10 & 98.32 & 96.70 & 0.00 & 98.32 & 96.70 & 0.00 & 97.92 & 96.90 & 1.03  \\        
& Derby & 94.90 & 90.35 & 0.00 & 94.90 & 90.35 & 0.00 & 95.59 & 92.54 & 1.10 & 94.92 & 90.46 & 0.11 & 95.22 & 91.56 & 0.77 & 95.35 & 94.41 & 3.62  \\
\rowcolor{gray!30} \cellcolor{white} & \textbf{Average} & 98.19 & 96.52 & 0.00 & 98.19 & 96.52 & 0.00 & \cellcolor{Green!60}98.32 & \cellcolor{Green!60}97.00 & \cellcolor{Green!60}0.28 & 98.19 & 96.54 & 0.04 & 98.27 & 96.78 & 0.15 & 98.14 & 97.43 & 1.12  \\
\hline
\multirow{6}*{\makecell[c]{BERT\\(uncased)}}  & Chromium & 99.77 & 99.54 & 0.00 & 99.77 & 99.54 & 0.00 & 99.77 & 99.55 & 0.01 & 99.77 & 99.54 & 0.00 & 99.77 & 99.54 & 0.00 & 99.78 & 99.63 & 0.07  \\
& Wicket & 99.49 & 98.99 & 0.00 & 99.44 & 98.89 & 0.00 & 99.49 & 98.99 & 0.00 & 99.49 & 98.99 & 0.00 & 99.49 & 98.99 & 0.00 & 99.44 & 98.99 & 0.10  \\
& Ambari & 98.48 & 97.01 & 0.00 & 98.48 & 97.01 & 0.00 & 98.44 & 97.22 & 0.31 & 98.48 & 97.01 & 0.00 & 98.43 & 97.01 & 0.10 & 97.98 & 97.01 & 1.03  \\
& Camel & 98.32 & 96.70 & 0.00 & 98.32 & 96.70 & 0.00 & 98.32 & 96.80 & 0.10 & 98.32 & 96.70 & 0.00 & 98.32 & 96.70 & 0.00 & 98.12 & 96.90 & 0.62  \\
& Derby & 94.90 & 90.35 & 0.00 & 94.90 & 90.35 & 0.00 & 95.65 & 92.65 & 1.10 & 94.98 & 90.57 & 0.11 & 95.49 & 91.89 & 0.55 & 95.53 & 94.52 & 3.40  \\
\rowcolor{gray!30} \cellcolor{white} & \textbf{Average} & 98.19 & 96.52 & 0.00 & 98.18 & 96.50 & 0.00 & \cellcolor{Green!60}98.33 & \cellcolor{Green!60}97.04 & \cellcolor{Green!60}0.30 & 98.21 & 96.56 & 0.02 & \cellcolor{Green!35}98.30 & \cellcolor{Green!35}96.83 & \cellcolor{Green!35}0.13 & 98.17 & 97.41 & 1.04  \\
\hline
\multirow{6}*{\makecell[c]{RoBERTa}}  & Chromium & 99.77 & 99.54 & 0.00 & 99.77 & 99.54 & 0.00 & 99.77 & 99.54 & 0.00 & 99.77 & 99.54 & 0.00 & 99.77 & 99.54 & 0.00 & 99.79 & 99.71 & 0.13  \\
& Wicket & 99.49 & 98.99 & 0.00 & 99.49 & 98.99 & 0.00 & 99.49 & 98.99 & 0.00 & 99.49 & 98.99 & 0.00 & 99.49 & 98.99 & 0.00 & 99.44 & 98.99 & 0.10  \\
& Ambari & 98.48 & 97.01 & 0.00 & 98.48 & 97.01 & 0.00 & 98.48 & 97.01 & 0.00 & 98.48 & 97.01 & 0.00 & 98.44 & 97.12 & 0.21 & 98.14 & 97.22 & 0.93  \\
& Camel & 98.32 & 96.70 & 0.00 & 98.32 & 96.70 & 0.00 & 98.32 & 96.70 & 0.00 & 98.32 & 96.70 & 0.00 & 98.27 & 96.70 & 0.10 & 98.17 & 96.90 & 0.52  \\
& Derby & 94.90 & 90.35 & 0.00 & 94.90 & 90.35 & 0.00 & 95.29 & 91.23 & 0.22 & 94.90 & 90.35 & 0.00 & 94.81 & 91.01 & 0.99 & 95.48 & 94.63 & 3.62  \\
\rowcolor{gray!30} \cellcolor{white} & \textbf{Average} & 98.19 & 96.52 & 0.00 & 98.19 & 96.52 & 0.00 & 98.27 & 96.69 & 0.04 & 98.19 & 96.52 & 0.00 & 98.16 & 96.67 & 0.26 & 98.20 & 97.49 & 1.06  \\
\hline
\multirow{6}*{\makecell[c]{distilRoBERTa}}  & Chromium & 99.77 & 99.54 & 0.00 & 99.77 & 99.54 & 0.00 & 99.77 & 99.54 & 0.00 & 99.77 & 99.54 & 0.00 & 99.77 & 99.54 & 0.00 & 99.78 & 99.67 & 0.10  \\
& Wicket & 99.49 & 98.99 & 0.00 & 99.49 & 98.99 & 0.00 & 99.49 & 98.99 & 0.00 & 99.49 & 98.99 & 0.00 & 99.49 & 98.99 & 0.00 & 99.49 & 98.99 & 0.00  \\
& Ambari & 98.48 & 97.01 & 0.00 & 98.48 & 97.01 & 0.00 & 98.48 & 97.01 & 0.00 & 98.48 & 97.01 & 0.00 & 98.59 & 97.22 & 0.00 & 97.98 & 97.12 & 1.13  \\
& Camel & 98.32 & 96.70 & 0.00 & 98.32 & 96.70 & 0.00 & 98.32 & 96.70 & 0.00 & 98.32 & 96.70 & 0.00 & 98.32 & 96.70 & 0.00 & 97.91 & 96.80 & 0.93  \\
& Derby & 94.90 & 90.35 & 0.00 & 94.90 & 90.35 & 0.00 & 95.15 & 90.90 & 0.11 & 94.90 & 90.35 & 0.00 & 95.08 & 91.23 & 0.66 & 95.80 & 94.74 & 3.07  \\
\rowcolor{gray!30} \cellcolor{white} & \textbf{Average} & 98.19 & 96.52 & 0.00 & 98.19 & 96.52 & 0.00 & 98.24 & 96.63 & 0.02 & 98.19 & 96.52 & 0.00 & 98.25 & 96.74 & 0.13 & 98.19 & 97.46 & 1.05  \\
\hline
\multirow{6}*{\makecell[c]{ALBERT}}  & Chromium & 99.77 & 99.54 & 0.00 & 99.77 & 99.54 & 0.00 & 99.78 & 99.60 & 0.03 & 99.77 & 99.54 & 0.00 & 99.77 & 99.54 & 0.00 & 99.77 & 99.62 & 0.09  \\
& Wicket & 99.49 & 98.99 & 0.00 & 99.49 & 98.99 & 0.00 & 99.44 & 98.99 & 0.10 & 99.49 & 98.99 & 0.00 & 99.49 & 98.99 & 0.00 & 99.44 & 98.99 & 0.10  \\
& Ambari & 98.48 & 97.01 & 0.00 & 98.48 & 97.01 & 0.00 & 98.38 & 97.12 & 0.31 & 98.48 & 97.01 & 0.00 & 98.48 & 97.01 & 0.00 & 98.08 & 97.12 & 0.93  \\
& Camel & 98.32 & 96.70 & 0.00 & 98.32 & 96.70 & 0.00 & 97.81 & 96.70 & 1.03 & 98.32 & 96.70 & 0.00 & 98.32 & 96.70 & 0.00 & 97.51 & 96.70 & 1.65  \\
& Derby & 94.90 & 90.35 & 0.00 & 94.90 & 90.35 & 0.00 & 94.98 & 92.76 & 2.63 & 94.90 & 90.35 & 0.00 & 94.99 & 91.23 & 0.88 & 94.25 & 93.09 & 4.49  \\
\rowcolor{gray!30} \cellcolor{white} & \textbf{Average} & 98.19 & 96.52 & 0.00 & 98.19 & 96.52 & 0.00 & 98.08 & 97.03 & 0.82 & 98.19 & 96.52 & 0.00 & 98.21 & 96.69 & 0.18 & 97.81 & 97.10 & 1.45  \\
\hline
\multirow{6}*{\makecell[c]{distilBERT\\(cased)}}  & Chromium & 99.77 & 99.54 & 0.00 & 99.77 & 99.54 & 0.00 & 99.77 & 99.54 & 0.00 & 99.77 & 99.54 & 0.00 & 99.77 & 99.54 & 0.00 & 99.75 & 99.64 & 0.14  \\
& Wicket & 99.49 & 98.99 & 0.00 & 99.49 & 98.99 & 0.00 & 99.49 & 98.99 & 0.00 & 99.49 & 98.99 & 0.00 & 99.49 & 98.99 & 0.00 & 99.39 & 98.99 & 0.20  \\
& Ambari & 98.48 & 97.01 & 0.00 & 98.48 & 97.01 & 0.00 & 98.43 & 97.01 & 0.10 & 98.43 & 97.01 & 0.10 & 98.48 & 97.01 & 0.00 & 98.08 & 97.01 & 0.82  \\
& Camel & 98.32 & 96.70 & 0.00 & 98.32 & 96.70 & 0.00 & 98.22 & 96.70 & 0.21 & 98.32 & 96.70 & 0.00 & 98.32 & 96.70 & 0.00 & 97.87 & 97.01 & 1.24  \\
& Derby & 94.90 & 90.35 & 0.00 & 94.90 & 90.35 & 0.00 & 95.86 & 92.65 & 0.66 & 94.86 & 90.35 & 0.11 & 95.15 & 91.34 & 0.66 & 95.34 & 93.86 & 3.07  \\
\rowcolor{gray!30} \cellcolor{white} & \textbf{Average} & 98.19 & 96.52 & 0.00 & 98.19 & 96.52 & 0.00 & \cellcolor{Green!35}98.35 & \cellcolor{Green!35}96.98 & \cellcolor{Green!35}0.19 & 98.17 & 96.52 & 0.04 & 98.24 & 96.72 & 0.13 & 98.09 & 97.30 & 1.09  \\
\hline
\multirow{6}*{\makecell[c]{distilBERT\\(uncased)}}  & Chromium & 99.77 & 99.54 & 0.00 & 99.77 & 99.54 & 0.00 & 99.77 & 99.54 & 0.00 & 99.77 & 99.54 & 0.00 & 99.77 & 99.54 & 0.00 & 99.76 & 99.67 & 0.16  \\
& Wicket & 99.49 & 98.99 & 0.00 & 99.44 & 98.89 & 0.00 & 99.49 & 98.99 & 0.00 & 99.49 & 98.99 & 0.00 & 99.49 & 98.99 & 0.00 & 99.44 & 98.99 & 0.10  \\
& Ambari & 98.48 & 97.01 & 0.00 & 98.48 & 97.01 & 0.00 & 98.43 & 97.12 & 0.21 & 98.43 & 97.01 & 0.10 & 98.43 & 97.01 & 0.10 & 97.88 & 97.12 & 1.34  \\
& Camel & 98.32 & 96.70 & 0.00 & 98.32 & 96.70 & 0.00 & 98.27 & 96.70 & 0.10 & 98.32 & 96.70 & 0.00 & 98.32 & 96.70 & 0.00 & 98.02 & 96.70 & 0.62  \\
& Derby & 94.90 & 90.35 & 0.00 & 94.90 & 90.35 & 0.00 & 95.80 & 92.65 & 0.77 & 94.81 & 90.35 & 0.22 & 95.11 & 91.56 & 0.99 & 95.04 & 94.63 & 4.50  \\
\rowcolor{gray!30} \cellcolor{white} & \textbf{Average} & 98.19 & 96.52 & 0.00 & 98.18 & 96.50 & 0.00 & \cellcolor{Green!80}\textbf{98.35} & \cellcolor{Green!80}\textbf{97.00} & \cellcolor{Green!80}\textbf{0.22} & 98.16 & 96.52 & 0.06 & 98.22 & 96.76 & 0.22 & 98.03 & 97.42 & 1.34  \\
\hline
\multirow{6}*{\makecell[c]{ELECTRA}}  & Chromium & 99.77 & 99.54 & 0.00 & 99.77 & 99.54 & 0.00 & 99.77 & 99.54 & 0.00 & 99.77 & 99.54 & 0.00 & 99.77 & 99.54 & 0.00 & 99.77 & 99.66 & 0.11  \\
& Wicket & 99.49 & 98.99 & 0.00 & 99.24 & 98.48 & 0.00 & 99.49 & 98.99 & 0.00 & 99.49 & 98.99 & 0.00 & 99.49 & 98.99 & 0.00 & 99.34 & 98.99 & 0.30  \\
& Ambari & 98.48 & 97.01 & 0.00 & 98.11 & 96.29 & 0.00 & 98.48 & 97.01 & 0.00 & 98.48 & 97.01 & 0.00 & 98.48 & 97.01 & 0.00 & 98.33 & 97.01 & 0.31  \\
& Camel & 98.32 & 96.70 & 0.00 & 98.10 & 96.28 & 0.00 & 98.32 & 96.70 & 0.00 & 98.32 & 96.70 & 0.00 & 98.32 & 96.70 & 0.00 & 98.17 & 96.80 & 0.41  \\
& Derby & 94.90 & 90.35 & 0.00 & 94.90 & 90.35 & 0.00 & 94.90 & 90.35 & 0.00 & 94.85 & 90.35 & 0.11 & 94.63 & 90.57 & 0.88 & 94.54 & 91.67 & 2.30  \\
\rowcolor{gray!30} \cellcolor{white} & \textbf{Average} & 98.19 & 96.52 & 0.00 & 98.02 & 96.19 & 0.00 & 98.19 & 96.52 & 0.00 & 98.18 & 96.52 & 0.02 & 98.14 & 96.56 & 0.18 & 98.03 & 96.83 & 0.69  \\
\Xhline{1pt}
\end{tabular}
}
\end{table*}

As shown in Table \ref{tbl5}, we select 8 language models and 6 machine learning algorithms for experiments, resulting in a total of $8\times6=48$ sets of results. Each set of results also includes 5 data sets and three indicators. Our experimental results are too large. To express the experimental results more clearly, we take the average of each set of experimental data in data set units to obtain each set's representative results and measure which combination is more suitable for SEDAC. Observing the representative results, the \emph{g-measure} of almost all combinations is above 98\%, except for the combination of ALBERT and MLP. Among all combinations, the best five are BERT (cased) and logistic regression, BERT (uncased) and logistic regression, distilBERT (cased) and logistic regression, distilBERT (uncased) and logistic regression, and BERT (uncased) and KNN, respectively, with \emph{g-measure} exceeding 98.3\%. Among these five combinations, the top three combinations with the highest \emph{pd} are BERT (cased) and logistic regression, BERT (uncased) and logistic regression, and distilBERT (uncased) and logistic regression, with \emph{pd} exceeding 97\%. Among them, the combination of distilBERT (uncased) and logistic regression slightly outperforms other 
 two ones with a lower \emph{pf}.

\subsubsection{RQ3.How effective is the CVAE model in the data synthesizing step?}
\begin{table} 
	\centering
	\caption{Comparison of SEDAC with and without CVAE module in \emph{g-measure(\%)}, \emph{pd(\%)}, and \emph{pf(\%)}.}  
	\label{tbl6}
    \resizebox{1.0\linewidth}{!}{
	\begin{tabular}{ccccc}
		\Xhline{1pt}
		\multirow{2}*{Model} & \multirow{2}*{Project} & \multicolumn{3}{c}{Logistic Regression} \\ 
		\Xcline{3-5}{0.7pt}
		& & g-measure & pd & pf \\
		\hline 
		\multirow{6}*{\makecell[c]{distilBERT\\(uncased)}} & Chromium & 99.77 & 99.54 & 0.00  \\
            & Wicket & 99.49 & 98.99 & 0.00   \\
            & Ambari & 98.43 & 97.12 & 0.21  \\
            & Camel & 98.27 & 96.70 & 0.10  \\
            & Derby & 95.80 & 92.65 & 0.77  \\
            \rowcolor{gray!30} \cellcolor{white} & \textbf{Average} & 98.35 & 97.00 & 0.22  \\
            \hline
            \multirow{6}*{\makecell[c]{distilBERT\\(uncased)\\\textit{w/o CVAE}}} & Chromium & 1.03 & 0.53 & 0.00  \\
            & Wicket & 0.00 & 0.00 & 0.00   \\
            & Ambari & 0.00 & 0.00 & 0.00  \\
            & Camel & 0.00 & 0.00 & 0.10  \\
            & Derby & 40.32 & 26.08 & 0.99  \\
            \rowcolor{gray!30} \cellcolor{white} & \textbf{Average} & 8.27(\textcolor{Green}{$\downarrow$91.59\%}) & 5.32(\textcolor{Green}{$\downarrow$94.51\%}) & 0.22  \\
            \Xhline{1pt}
        \end{tabular}
    }
\end{table}

\paragraph{Approach}
The CVAE model is the core of SEDAC. To evaluate the effectiveness of the CVAE model in SEDAC, we removed the CVAE module and classified the original imbalanced data set. 
\paragraph{Results}
The results are shown in Table \ref{tbl6}. For Chromium, \emph{g-measure} is only 1.03\%, and \emph{pd} is only 0.53\%. For Wicket, Ambari, and Camel, both \emph{g-measure} and \emph{pd} become 0\%. For Derby, although \emph{g-measure} is still 40.32\%, compared with the original result of 95.80\%, it still drops by nearly 60\%, and \emph{pd} drops by about 70\%. On average, the degradation performance of SEDAC without the CVAE module is reflected in the decline rate of 91.59\% in \emph{g-measure} and the decline rate of 94.51\% in \emph{pd}. In summary, the performance of CVAE in the data synthesis step is very effective.

\section{Discussion}

\subsection{Reasons for Improvement}

\subsubsection{BERT Variant} DistilBERT (uncased), as a variant of BERT, well inherits the function of capturing contextual information of the BERT model and provides an adequate representation for SEDAC. Although other BERT variants have also demonstrated their excellent embedding capabilities, distilBERT slightly surpasses other variants' average results based on their performance when used with different machine learning algorithms.

\subsubsection{The CVAE Model} The CVAE model provides an adequate data augmentation method for SEDAC. This data augmentation method is for input data with labels. It generates rare SBRs for imbalanced data sets, thereby constructing a balanced data set and improving the performance of SBR identification. Compared with other augmentation methods that do not use labels, the characteristics of the data generated by the CVAE model are more similar to the original data. Moreover, its encoder-decoder architecture is well suited for processing sequential data, such as converted bug report vectors in our approach.

\subsubsection{Balanced Data} Completely balanced data is the fundamental reason for the excellent performance of various indicators in the SBR identification stage. It is also SEDAC's core competitiveness that outperforms other baselines. Previous research methods, whether filtering NSBRs or synthesizing SBRs, did not achieve a proper balance between them. Since the proportion of SBRs in the project is extremely low, filtering NSBRs is a temporary solution rather than the root cause. It leads to information loss, and the filtered data set still suffers from data imbalance. Although our method injects new data into the original data set, our classification effect is excellent because the generated security bug report vectors are very similar but different from the original ones,  regardless of which machine learning algorithm is used.

\subsection{Quality of Synthesized Data}
\begin{table*}
\centering
\caption{Duplicated number of vectors generated by SEDAC. Original(\%) is the proportion of duplicated bug report vectors in the original data. Synthesized(\%) is the proportion of duplicated bug report vectors in the generated data.}
\label{tbl7}
\resizebox{0.7\linewidth}{!}{
\begin{tabular}{cccccc}
\toprule
 & Duplicated & Original & Original(\%) & Synthesized & Synthesized(\%)  \\
\midrule
Chromium & 121 & 41940 & 0.29 & 41556 & 0.29 \\
Wicket & 1 & 1000 & 0.10 & 980 & 0.10 \\
Ambari & 1 & 1000 & 0.10 & 942 & 0.11 \\
Camel & 0 & 1000 & 0.00 & 936 & 0.00 \\
Derby & 0 & 1000 & 0.00 & 824 & 0.00 \\
\bottomrule
\end{tabular}
}
\end{table*}

Although SEDAC performs well, since the amount of generated data is close to the original one, we need to consider whether there is a large amount of the same data, resulting in a perfect classification effect. Therefore, we perform duplication-checking on the synthesized data and obtain the results in Table \ref{tbl7}. \emph{Original} refers to the number of bug reports in the original datasets, and \emph{Original(\%)} is the proportion of duplicated bug report vectors in it. \emph{Synthesized} is the number of synthesized bug report vectors and \emph{Synthesized(\%)} is the proportion of duplicated ones in it. We can see that in the larger project Chromium, there are 121 duplicate data, accounting for only 0.29\% in both the original and the generated data. In Wicket and Ambari, there is one piece of duplicate data each, accounting for about 0.1\% of the original and generated data. In Camel and Derby, there is no duplicate data. Since the proportion of these repeated data is very low, we can ignore their impact on the SBR identification stage.

\subsection{Threats to Validity}
\subsubsection{Validity of Raw Data} To make a fair comparison with other baselines, we have not updated these five datasets from Chromium and Apache projects. They were all collected and organized several years ago, and newer data sets may positively impact our framework. In addition, the number of SBRs in the four Apache projects is only 1,000. Although they represent the performance of our framework on small data sets, rich data may be more conducive to model training. In addition, the labels of these organized data sets are all manually labeled, which may need to be corrected, thus affecting the effectiveness of our framework.

\subsubsection{Generalizability} Our training and testing process are only implemented on Chromium, Wicket, Ambari, Camel, and Derby projects. However, we encounter more projects in different fields in the real world. On the one hand, whether our approach is also practical for other projects is still being determined. On the other hand, the number of real-world data with labels that can be used for model training may be minimal. Therefore, there may be threats to the generalization ability of our model. More projects should be involved in our experiments.

\section{Conclusion}
This paper proposes SEDAC, a CVAE-based data augmentation method for identifying security bug reports. SEDAC is a novel method that can solve the data imbalance problem in SBR identification by generating SBRs. Our key idea is to (1) use the distilBERT model to convert summary and description in bug reports into bug report vectors, (2) use the bug report vectors and corresponding labels as input data and condition vectors to train the CVAE model, (3) separate the decoder in the CVAE model and use it as a generator to generate SBR vectors, and (4) combine the generated SBR vectors with the original vectors to form a balanced data set for training SBR classifier. We evaluated SEDAC on the chromium project, and four Apache projects, and the experimental results proved that SEDAC outperforms the baselines. In the future, we intend to apply SEDAC to more real-world projects.

\bibliography{reference}
\bibliographystyle{IEEEtran}

\end{document}